\documentstyle[12pt]{article}

\begin{document}
\begin{center}
{\Large How Good Can We Get?} \\
{\large Using mathematical models to predict the future of athletics}\\

\vspace{5mm}
J. R. Mureika \\
{\it Department of Computer Science \\
University of Southern California \\
Los Angeles, CA~~90089-2520}
\end{center}
 
There are numerous rewards which world class athletes covet in their quests
for glory, be they National titles, World Champion status or Olympic Gold.
Perhaps one of the top honors, though, is that of the World Record: the
symbol and defining mark that the highest level in one's event has been
achieved.\\

World records have risen and fallen throughout the history of Track and
Field, and the past 2 years have certainly been no exception. In fact, a
significant portion of the men's Track marks have been re-written since
1993. Of current interest are the progressions of the middle distance kings,
Hicham El Guerrouj in the 1500m and mile, the ongoing battle for supremacy
between Daniel Komen and Haile Gebrselassie, and the unforgettable,
unquestionable dominance of Wilson Kipketer in the 800m. At the time of this
writing, Maria Mutola has set a new indoor 800m WR, while Komen and Geb are
still waging their war of attrition under the roof. The sprint records are
yet again on the verge of being knocked down a few notches. Marion Jones has
her sights set on Irina Privalova's 60m mark, and is mumbling about 10.5s
clockings this summer. Similarly, Maurice Greene's ground-breaking in the
indoor sprints, as well as his February 28th 9.99s dash Down Under (the
first ever sub-10s clocking on Aussie soil, with a -0.6 m/s wind, no less!),
seem to be indicative of the shape of things to come.\\

On this note, we have to ask ourselves: "What {\it are} the shape of things to
come?". A record is set, a record falls. Can this process go on
indefinitely? It seems obvious to answer 'no' to such a question, but we are
then posed with a corollary: exactly how good can we get?\\

The answer to this question has been the subject of various research works
over the decades (see the references), and in the following few paragraphs,
I'll sum up the findings of one paper which I have found particularly
fascintating. This interesting and insightful work is an article by Fran\c{c}ois
P\'{e}ronnet and Guy Thibault, then of McGill University, researched in 1987 and
published in 1989. They posit a mathematical model which reproduces an
athlete's power output over a given distance of running, using variables
representing various metabolic energy-yielding processes. This is in turn
based on a model developed by Arthur Hill way back in the 1920s.\\

\noindent{\bf The Aerobic and Anaerobic Physiological Variables} \\

Peronnet and Thibault studied variations in the following quantities in
athletic performances since the early part of the century: capacity of
anaerobic metabolism (raw strength), maximal aerobic power (peak endurance),
and the reduction in peak aerobic power with increasing race duration
(roughly overall endurance). They found that while the first two of these
increase in an essentially linear fashion over the years for all events, the
latter of the three remained effectively constant. That is, while
improvements in training techniques have helped to improve anaerobic and
aerobic capacity, the rate at which maximal exertion drops has not changed
much.\\

By studying these trends in the energy-yielding processes, Peronnet and
Thibault were able to pin down possible future performances for both men and
women. In particular, since the reduction in peak aerobic power does not
change, by placing reasonable "plateaus" on the first two aerobic and
anaerobic variables for human beings, the "ultimate" performances in track
could be guestimated.\\

\noindent{\bf Comments on altitude/enhanced performances and initial data} \\

The predictions in Tables 3-5 were made based in part on WRs dating before
and up to 1987, when the study was performed. Tables 1 and 2 present the
marks as of 1987, and provide a comparison with today's equivalents. Earlier
models which have attempted to predict future performances incorporated
these altitude performances, which served to skew the results, since the
``aided'' sprint marks from 1975 actually surpassed the 1987 ``sea-level''
best performances. While I have included altitude performances in the
records of Tables 1 and 2, Peronnet and Thibault used "sea-level WRs" in
their calculations. For example, Mennea's 1979 WR of 19.72s in the 200m was
run in Mexico City (at altitude), while the "sea-level" record was Carl
Lewis' 19.75s in 1983 (Indianapolis). This is also the case with Lee Evans'
43.86A from the 1968 Olympics, compared to the next fastest time, Butch
Reynold's 44.10s (Columbus, OH, 1987).\\

Similarly, the 9.83s 100m WR used in the study (it had not yet been
stricken) was in retrospect 10 years ahead of it's time, based on the
``natural'' progression of the short sprint (albeit, some of today's
training aids may be somewhat questionable). In fact, had it not been for
Carl Lewis' 9.93s clocking behind Ben's 1987 WC race (and various other
stricken Ben marks), Mel Lattany's (USA) 9.96s from Athens, GA, in 1984
should have been the mark used for initial data. This might cause the
predicted 100m times to be slightly overestimated.\\

\noindent{\bf The Model Predictions: Then and Now}\\

Tables 3 and 4 present the projected WR marks for men and women through the
years 2000, 2028, and 2040, as calculated by Peronnet and Thibault. In
physics, it's always interesting to see how good (or how bad!) one's model
actually is when compared to the real life situation it's trying to emulate.
When the original paper was written, the turn of the century was at least a
dozen years down the road. In contrast, today we have a much better idea of
how the record books may look in 2 years time!\\

The most obvious discrepancies of their model arise in the sprints, most
likely due to inaccuracies in pervious records (hand times, timing accuracy,
or... miscellaneous). Both the women's 100m and 200m predictions for 2000
were surpassed a year after they were made by one Florence Griffith-Joyner.
Although, the alleged 10.49s WR is subject to much debate (at least by those
of us who study wind effects on sprints!). In the event that this mark was
wind-aided, the current 100m WR should probably be in the low 10.6s range,
but still be awarded to Flo-Jo (10.61s, Indianapolis, 1988).\\

The men's 400m estimate of 43.44s was also surpassed in 1988 by Butch
Reynolds. Despite the 9.83s mark used as initial data for the 100m, the
predicted record of 9.74s is surprisingly close to becoming reality (perhaps
even as early as this summer?). In fact, this raises an interesting
question. If such a clocking were to come about shortly, then based on the
previous section's discussion (i.e. 9.83s), is it safe to assume that it is
"natural"?...\\

We can't discount the possibility of a 1:39.88 800m sometime soon, either.
Rumor has it that the name Kipketer could figure prominently. And, last but
certainly not least, in their calculations for the 200m WR, no initial data
could have let Peronnet and Thibault predict Michael Johnson! Based on their
study, a 19.32s deuce should not have occurred until about the year 2015.
The only logical explanation for this, of course, is that Johnson was sent
here from the future for some reason we cannot yet understand, but whose
purpose will soon be evident....\\

Performances further down the road are a bit more subjective, and we can't
for sure tell whether or not they are vast overestimates, or naive
underestimates. An 18.92s 200m is foreseen sometime around 2040. Others have
forecast an 18.97s sprint as early as 2004 [reference 2]. In fact, in my own
work on sprint curve-running [ref. 3], I projected that a sub-19s clocking
wasn't necessarily that far out of reach, based on Johnson's WR. So, whether
or not these are somewhat over-optimistic is open to debate.\\

After breaking the 4-minute mile, Roger Bannister ventured to say that a
3:30 clocking could be realizable by the year 1990. As much as El Guerrouj
might be interested, it doesn't quite seem possible at this time. Peronnet
and Thibault set this mark as attainable just shy of the mid-21st century.
Although, the 3:41.96 is slightly more accessible. The 12-minute barrier in
the 5000m was cited to not be challenged until the 2030s, but a sub
14-minute 5k for the women might come much sooner. Likewise, a sub-2 hour
marathon is predicted by the year 2028, as is the breaking of the 4-minute
mile barrier for the women.\\

Note that the predictions of the women's performances extends only to the
year 2033, as opposed to 2040 for the men. This arises due to the less
accurate and less numerous data obtained for women's WRs throughout the
century. As a result, the authors contend that the predictions for women's
records are probably underestimated in certain cases.\\

\noindent{\bf To the Limit...}\\

Many suggest that we cannot keep improving indefinitely. Certainly, there
are physiological (not to mention physical) limitations on the human body
that would prevent this. Based on the evolution of their model, Peronnet and
Thibault took a stab at what could be the ultimate performances in
athletics. These are presented in Table 5, and put forth a number of
interesting speculations. For the men, there will be no sub-9s 100m; we
shall never clock a 1:29 800m; there will always be an insurmountable
3-minute barrier for the mile. The 5000m will go sub-12, but will barely
drop another minute, and the marathon record will max out a mere 18 minutes
below its current value.\\

Many of these predictions seem to suggest that athletically, women may not
progress beyond the present level of the men. As mentioned earlier, Peronnet
and Thibault stress that due to insufficient data relative to the men's
marks, the predictions for women's performances may be underestimates, and
that their performances may actually approach the limiting values for men.
So, contrary to their findings in Table 5, we may someday be talking about
the first sub-10s women's sprint or 1:50:00 marathon.\\

Thus ends the gaze through the looking glass. Like all mathematical models,
the findings of this study should be taken with a grain of salt. They're not
meant to accurately represent the way things will be, but rather the way
they might be, based on present data. Fundamentally, we should not place
limits on ourselves. With them, we can only run faster in our dreams,
forever constrained by infinitely high walls. Without them, we can chase our
dreams into reality, and the impenetrable barriers are ours to break.\\

\vskip .25 cm
\noindent
 
\noindent
{\bf Acknowledgements}

I thank Fran\c{c}ois P\'{e}ronnet for proof-reading the article and offering 
comments. Dr. P\'{e}ronnet is now with the D\'{e}partement
d'\'{e}ducation physique \`{a} l'Universit\'{e} de Montr\'{e}al 
(peronnet@ere.umontreal.ca), and is currently on leave at 
l'Universit\'{e} Joseph Fourier in Grenoble, France. I also thank Natasha Bayus 
of the University of Southern California for insightful and motivational
discussions on the subject.

\pagebreak

\begin{table}[t]
\begin{center}
{\begin{tabular}{|l|| l| l|}\hline
 &   {\bf 1987 WR}               &           {\bf 1997 WR} \\ \hline
100m &  9.83s (Ben Johnson, CAN)       &  9.84s (Donovan Bailey, CAN)\\
200m &  19.72A (Pietro Mennea, ITA)    &  19.32s (Michael Johnson, USA)\\
400m &  43.86A (Lee Evans, USA)        &  43.39s (Butch Reynolds, USA)\\
800m &  1:41.73 (Sebastian Coe, GBR)   &  1:41.11 (Wilson Kipkepter, DEN)\\
1000m&  2:12.18 (Sebastian Coe, GBR)   &  2:12.18 (Sebastian Coe, GBR)\\
1500m&  3:29.46 (Said Aouita, MOR)     &  3:27.37 (Noureddine Morceli, ALG)\\
Mile &  3:46.32 (Steve Cram, GBR)      &  3:44.39 (Noureddine Morceli, ALG)\\
2000m&  4:50.81 (Said Aouita, MOR)     &  4:47.88 (Noureddine Morceli, ALG)\\
3000m&  7:32.1 (Henry Rono, KEN)       &  7:20.67 (Daniel Komen, KEN)\\
5000m&  12:58.39 (Said Aouita, MOR)    &  12:39.74 (Daniel Komen, KEN)\\
10000m& 27:13.81 (Fernando Mamede, POR)&  26:27.85 (Paul Tergat, KEN)\\
Mar. &  2:07:12 (Carlos Lopes, POR)    &  2:06:50 (Belayneh Dinsamo, ETH)\\ \hline
\end{tabular}}
\end{center}
\caption{Men's World Records}
\label{table1}
\end{table}

\begin{table}[t]
\begin{center}
{\begin{tabular}{|l|| l| l|}\hline
          &   {\bf 1987 WR}               &           {\bf 1997 WR} \\ \hline
100m  & 10.76s (Evelyn Ashford, USA)   &  10.49s (Florence Griffith-Joyner, USA)\\
200m  & 21.71s (Heike Drechsler, DDR)  &  21.34s (Flo. Griffith-Joyner, USA)\\
400m  & 47.60s (Marita Koch, DDR)      &  47.60s (Marita Koch, DDR)\\
800m  & 1:53.28 (Jarmila Kratochvilova, TCH)&  1:53.28 (Jarmila Kratochvilova, TCH)\\
1000m & 2:30.6 (Tatyana Provodikina,SOV)& 2:28.98 (Svetlana Masterkova, RUS)\\
1500m & 3:52.47 (Tatyana Kazankina, SOV)& 3:50.46 (Qu Yanxia, CHN)\\
Mile  & 4:15.8 (Natalja Artemova, SOV)  & 4:12.56 (Svetlana Masterkova, RUS)\\
2000m & 5:28.69 (Maricica Puica, ROM)   & 5:25.36 (Sonia O'Sullivan, IRE)\\
3000m & 8:22.62 (Tatyana Kazankina, SOV)& 8:06.11 (Wang Yungxia, CHN)\\
5000m & 14:37.33 (Ingrid Kristiansen, NOR)   & 14:28.09 (Jiang Bo, CHN)\\
10000m& 30:13.74 (Ingrid Kristiansen,  NOR)  & 29:31.78 (Wang Yungxia, CHN)\\
Mar.  & 2:21:06 (Ingrid Kristiansen,NOR)& 2:21:06 (Ingrid Kristiansen, NOR)\\ \hline
\end{tabular}}
\end{center}
\caption{Women's World Records}
\label{table2}
\end{table}

\begin{table}[t]
\begin{center}
{\begin{tabular}{|l| l l l|}\hline
           &  {\bf 2000}     &     {\bf 2028}    &    {\bf  2040} \\ \hline
100m      &    9.74s      &        9.57s       &       9.49s\\
200m      &    19.53s     &        19.10s      &       18.92s\\
400m      &    43.44s     &        42.12s      &       41.59s\\
800m      &    1:39.88    &        1:36.18     &       1:34.71\\
1000m     &    2:09.72    &        2:04.81     &       2:02.86\\
1500m     &    3:25.45    &        3:17.45     &       3:14.27\\
Mile      &    3:41.96    &        3:33.29     &       3:29.84\\
2000m     &    4:45.15    &        4:33.89     &       4:29.41\\
3000m     &    7:22.54    &        7:03.91     &       6:56.87\\
5000m     &    12:42.72   &        12:09.39    &       11:56.19\\
10000m    &    26:43.63   &        25:32.27    &       25:04.01\\
Mar.      &    2:05:24    &        1:59:36     &       1:57:18\\ \hline
\end{tabular}}
\end{center}
\caption{Predicted future men's WR, based on 1987 records (from [1])}
\label{table3}
\end{table}

\begin{table}[t]
\begin{center}
{\begin{tabular}{|l| l l l|}\hline
           &  {\bf 2000}     &     {\bf 2028}    &    {\bf  2033} \\ \hline
100m       &   10.66s        &     10.46s        &     10.44s \\
200m       &   21.46s        &     20.95s        &     20.90s \\
400m       &   46.85s        &     45.34s        &     45.18s \\
800m       &   1:51.16       &     1:46.95       &     1:46.53 \\
1000m      &   2:27.74       &     2:22.03       &     2:21.45 \\
1500m      &   3:47.93       &     3:38.91       &     3:38.00 \\
Mile       &   4:10.79       &     4:00.83       &     3:59.82 \\
2000m      &   5:22.19       &     5:09.27       &     5:07.96 \\
3000m      &   8:11.98       &     7:50.61       &     7:48.46 \\
5000m      &   14:19.33      &     13:41.56      &     13:37.75 \\
10000m     &   29:38.41      &     28:19.04      &     28:11.04 \\
Mar.       &   2:18:43       &     2:12:20       &     2:11:41 \\ \hline
\end{tabular}}
\end{center}
\caption{Predicted future women's WR, based on 1987 records (from [1])}
\label{table4}
\end{table}

\begin{table}[t]
\begin{center}
{\begin{tabular}{|l|| l| l|}\hline
           &         {\bf Men}      &           {\bf Women} \\ \hline
100m       &         9.37s          &           10.15s \\
200m       &         18.32s         &           20.25s \\
400m       &         39.60s         &           44.71s \\
800m       &         1:30.86        &           1:42.71 \\
1000m      &         1:57.53        &           2:12.50 \\
1500m      &         3:04.27        &           3:26.95 \\
Mile       &         3:18.87        &           3:43.24 \\
2000m      &         4:11.06        &           4:41.48 \\
3000m      &         6:24.81        &           7:11.42 \\
5000m      &         11:11.61       &           12:33.36 \\
10000m     &         23:36.89       &           26:19.48 \\
Mar        &         1:48:25        &           2:00:33 \\ \hline
\end{tabular}}
\end{center}
\caption{Predicted ultimate performances for men and women (from [1])}
\label{table5}
\end{table}

\end{document}